\documentclass[twocolumn]{aastex62}

\usepackage{amsmath}
\usepackage{color}
\graphicspath{{./}{figures/}}
\usepackage{natbib}
\usepackage{enumitem}
\usepackage{float}
\usepackage{url}
\usepackage{color,soul}

\newcommand{\sersic}{S\'ersic}
\def\farcs{\hbox{$.\!\!^{\prime\prime}$}}

\journalinfo{The Astrophysical Journal Letters}
\submitjournal{ApJ Letters}

\shorttitle{The Globular Clusters in NGC5846-UDG1}
\shortauthors{Danieli et al.}

\begin{document}

\title{NGC5846-UDG1: A galaxy formed mostly by star formation in massive, extremely dense clumps of gas}

\correspondingauthor{Shany Danieli}
\email{sdanieli@astro.princeton.edu, shanyi1@gmail.com}

\author[0000-0002-1841-2252]{Shany Danieli}
\altaffiliation{NASA Hubble Fellow}
\affil{Department of Astrophysical Sciences, 4 Ivy Lane, Princeton University, Princeton, NJ 08544, USA \\}
\affil{Institute for Advanced Study, 1 Einstein Drive, Princeton, NJ 08540, USA \\}

\author[0000-0002-8282-9888]{Pieter van Dokkum}
\affiliation{Department of Astronomy, Yale University, New Haven, CT 06511, USA \\}

\author[0000-0003-2482-0049]{Sebastian Trujillo-Gomez}
\affiliation{Astronomisches Rechen-Institut, Zentrum für Astronomie der Universität Heidelberg, Monchhofstraße 12-14, D-69120 Heidelberg, Germany \\}

\author[0000-0002-8804-0212]{J.~M.~Diederik Kruijssen}
\affiliation{Astronomisches Rechen-Institut, Zentrum für Astronomie der Universität Heidelberg, Monchhofstraße 12-14, D-69120 Heidelberg, Germany \\}

\author[0000-0003-2473-0369]{Aaron J. Romanowsky}
\affiliation{Department of Physics and Astronomy, San Jos\'e State University, San Jose, CA 95192, USA\\}
\affiliation{University of California Observatories, 1156 High Street,
Santa Cruz, CA 95064, USA\\}

\author[0000-0002-5382-2898]{Scott Carlsten}
\affil{Department of Astrophysical Sciences, 4 Ivy Lane, Princeton University, Princeton, NJ 08544, USA \\}

\author[0000-0002-5120-1684]{Zili Shen}
\affiliation{Department of Astronomy, Yale University, New Haven, CT 06511, USA \\}

\author[0000-0001-9592-4190]{Jiaxuan Li}
\affil{Department of Astrophysical Sciences, 4 Ivy Lane, Princeton University, Princeton, NJ 08544, USA \\}

\author[0000-0002-4542-921X]{Roberto Abraham}
\affiliation{Department of Astronomy and Astrophysics, University of Toronto, Toronto ON, M5S 3H4, Canada\\}
\affiliation{Dunlap Institute for Astronomy and Astrophysics, University of Toronto, Toronto ON, M5S 3H4, Canada\\}

\author[0000-0002-9658-8763]{Jean Brodie}
\affiliation{Centre for Astrophysics \& Supercomputing, Swinburne University of Technology, Hawthorn VIC 3122, Australia\\}
\affiliation{University of California Observatories, 1156 High Street, Santa Cruz, CA 95064, USA\\}

\author[0000-0002-1590-8551]{Charlie Conroy}
\affiliation{Harvard-Smithsonian Center for Astrophysics, 60 Garden Street, Cambridge, MA, USA\\}

\author{Jonah S. Gannon}
\affiliation{Centre for Astrophysics \& Supercomputing, Swinburne University of Technology, Hawthorn VIC 3122, Australia\\}

\author[0000-0003-4970-2874]{Johnny Greco}
\affiliation{Center for Cosmology and AstroParticle Physics (CCAPP), The Ohio State University, Columbus, OH 43210, USA\\}

\begin{abstract}
It has been shown that ultra-diffuse galaxies (UDGs) have higher specific frequencies of globular clusters on average than other dwarf galaxies with similar luminosities. The UDG NGC5846-UDG1 is among the most extreme examples of globular cluster-rich galaxies found so far. Here we present new Hubble Space Telescope ({\it HST}) observations and analysis of this galaxy and its globular cluster system.  We find that NGC5846-UDG1 hosts $54 \pm 9$ globular clusters, three to four times more than any previously known galaxy with a similar luminosity, and higher than reported in previous studies. With a galaxy luminosity of $L_{V,\mathrm{gal}} \approx 6 \times 10^7\,{\rm L}_{\odot}$ ($M_\star \approx 1.2 \times 10^8\,{\rm M}_\odot$) and a total globular cluster luminosity of $L_{V,\mathrm{GCs}} \approx 7.6 \times 10^6\,{\rm L}_{\odot}$, we find that the clusters currently comprise $\sim 13 \%$ of the total light. Taking into account the effects of mass loss from clusters during their formation and throughout their lifetime, we infer that most of the stars in the galaxy likely formed in globular clusters, and very little to no ``normal'' low-density star formation occurred. This result implies that the most extreme conditions during early galaxy formation promoted star formation in massive and dense clumps, in contrast to the dispersed star formation observed in galaxies today.
\end{abstract}

\keywords{galaxies: photometry -- galaxies: dwarf -- galaxies: star clusters: general -- galaxies: formation -- galaxies: star formation -- galaxies: individual (NGC5846-UDG1)}

\section{Introduction}\label{sec:intro}
Star formation in the Milky Way typically proceeds in molecular clouds with sizes of $10-100$\,pc, leading to loose conglomerations of stars that slowly disperse within the Galaxy \citep{Kennicutt2012:Starformation}. Star formation can also produce compact, gravitationally bound systems \citep{Krumholz2019:clusters}, the most massive of which become long-lived globular clusters \citep{Kruijssen2014:GCs}. This mode of star formation is rare because it requires extreme gas pressures, $P/k>10^6~{\rm K}~{\rm cm^{-3}}$ \citep{Elmegreen1997:gcs, Kruijssen2015}, causing globular clusters to contain less than 0.5\,\% of the stars in most present-day galaxies \citep{Forbes2018:gcs}. 

Nonetheless, the specific frequency ($S_\mathrm{N}$), the total number of clusters per unit galaxy luminosity, can differ by a factor of $\sim 40-50$ between individual galaxies, with the largest specific frequencies observed at the very lowest or very highest luminosities \citep{Miller2007:gcs, Harris2013:gcs}. Alongside their extended sizes and low surface brightnesses, some ultra-diffuse galaxies (UDGs; \citealt{vD2015:udgs}) stand out by their elevated globular cluster abundances and the high specific frequencies compared to other galaxies with the same luminosities (\citealt{Peng2016:DF17, vD2017:gcs, Lim2020:NGVS} and references therein). One proposed explanation is that UDGs formed at earlier times than typical dwarf galaxies, in higher surface density environments typical of these epochs, giving rise to a larger fraction of their stellar mass formed in gravitationally bound clusters, (e.g., \citealt{TG2021:feedback, Carleton2021:gcs}). 

In particular, one such extremely globular cluster-rich UDG is NGC5846-UDG1. It was first cataloged in a photometric survey targeting the area surrounding the NGC5846 group with the Canada--France--Hawaii Telescope \citep{Mahdavi2005:udg1} and was recently re-identified in the VEGAS survey \citep{Forbes2019:udg1}, showing a collection of globular clusters at its center. Follow-up spectroscopic observations with the VLT multi-unit spectroscopic explorer (MUSE) confirmed that at least 11 of the globular clusters are associated with NGC5846-UDG1 based on their radial velocities \citep{Muller2020:muse}. The ground-based images lack the resolution to resolve the compact sources and therefore reliably determine the actual size of NGC5846-UDG1's globular cluster population. \citet{Muller2021:HST}\footnote{\citet{Muller2021:HST} refers to NGC5846-UDG1 as MATLAS-2019.} used a single orbit {\it HST}/ACS observations and identified $26 \pm 6$ globular clusters associated with NGC5846-UDG1.

In this Letter we present an analysis of NGC5846-UDG1 and its globular cluster system using new, deeper observations with {\it HST}/WFC3. We find a larger number of globular clusters than previous studies, and that cluster stars make up a remarkably high fraction of the total number of stars in the galaxy. We model NGC5846-UDG1's initial cluster mass function as a function of the birth galactic environment and discuss the implications of our results for different modes of star formation in the early Universe.

\begin{figure*}[t!]
{\centering
  \includegraphics[width=1.0\textwidth]{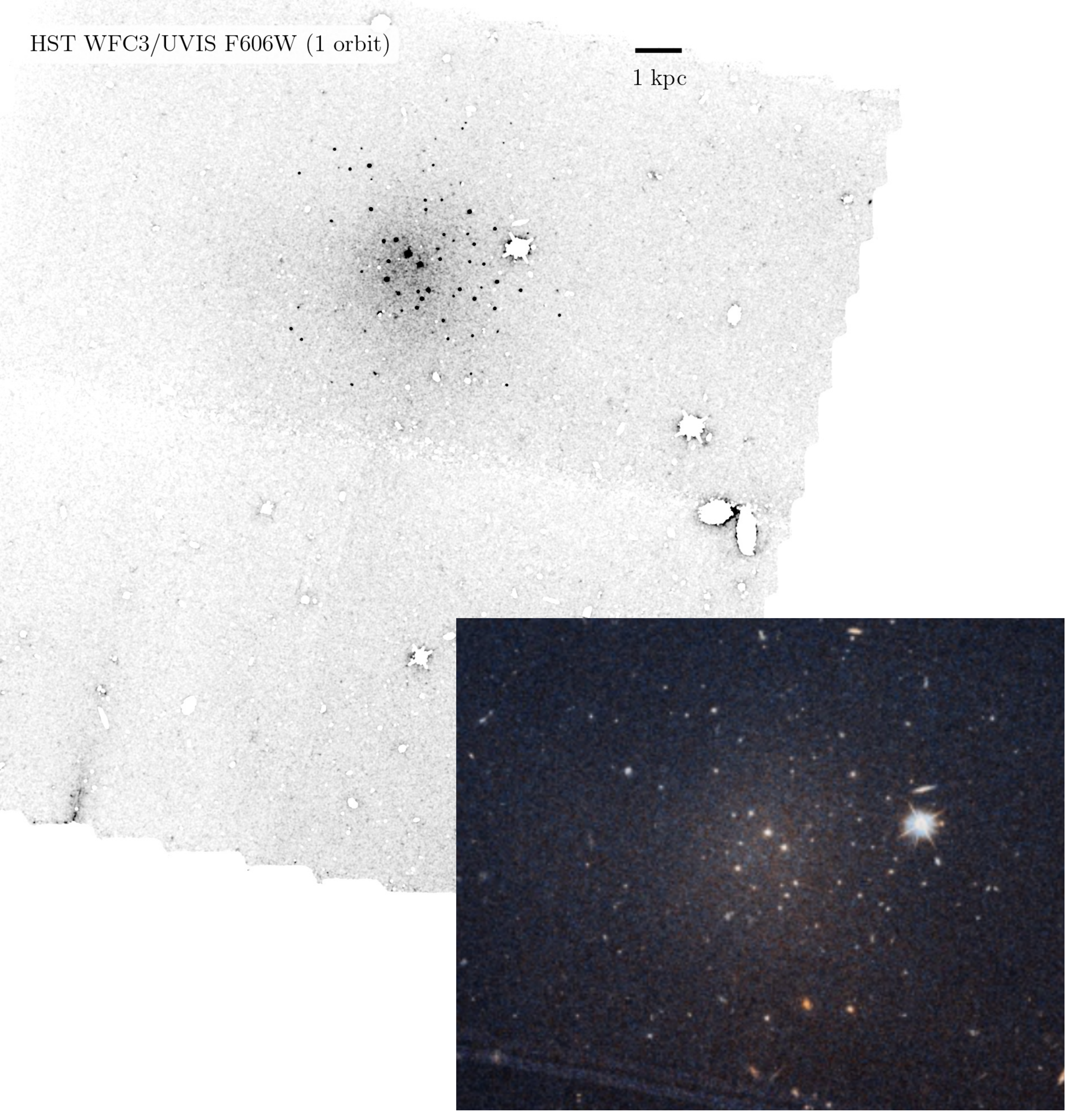}
  \caption{Upper image: $V$-band image taken with the Wide Field Camera 3 (WFC3) UVIS channel. All sources are masked except for the smooth diffuse light component of NGC5846-UDG1 and its globular cluster candidates. $\sim 13\%$ of all stars reside in NGC5846-UDG1's globular clusters and both components have similar spatial distributions. Lower right image: F606W$-$F475W combined color image. Globular clusters brighter than $M_\mathrm{F606W}=-7.6\,\mathrm{mag}$ have a median $g-V$ color of F475W$-$F606W$=0.4\,\mathrm{mag}$ and very little spread ($\sigma_{\mathrm{F475W}-\mathrm{F606W}}=0.03\,\mathrm{mag}$), suggesting similar ages and chemical compositions.}
  \label{fig:HSTimage}
}
\end{figure*}

\section{Observations and Analysis} \label{sec:observations}

\subsection{HST Imaging}
NGC5846-UDG1 was observed with WFC3/UVIS on 2020 December 27, in Cycle 28 (program GO-16284, PI: Danieli). The program focused on characterizing the numerous compact sources in the vicinity of NGC5846-UDG1 by obtaining deep images of the galaxy in two filters. Two orbits were obtained with a total exposure time of $2349\,\mathrm{sec}$ and $2360\,\mathrm{sec}$ in F475W and F606W, respectively. As the aim was to resolve individual globular clusters potentially associated with NGC5846-UDG1, the WFC3 camera was selected to exploit its improved resolution (0\farcs04/pixel) over the ACS camera. The drizzled \texttt{drc} images produced by the STScI standard pipeline were used in the analysis. The {\it HST} images were primarily used for the identification and characterization of the globular clusters in NGC5846-UDG1. The data were also used for isolating the low surface brightness component of the galaxy in the Dark Energy Camera Legacy Survey (DECaLS; \citealt{Dey2019:desi}) for deriving NGC5846-UDG1's structural parameters (see below).

A color image generated from the {\it HST} F606W and F475W images is shown in the bottom panel of Figure \ref{fig:HSTimage}. An abundant population of bright globular clusters is resolved with the {\it HST}/WFC3 camera's high resolution and can be seen clustered close to the center of the galaxy. A diffuse distribution of field stars is also detected. No other known galaxy has such a striking, dominant globular cluster population.

\begin{figure*}[t!]
{\centering
  \includegraphics[width=0.99\textwidth]{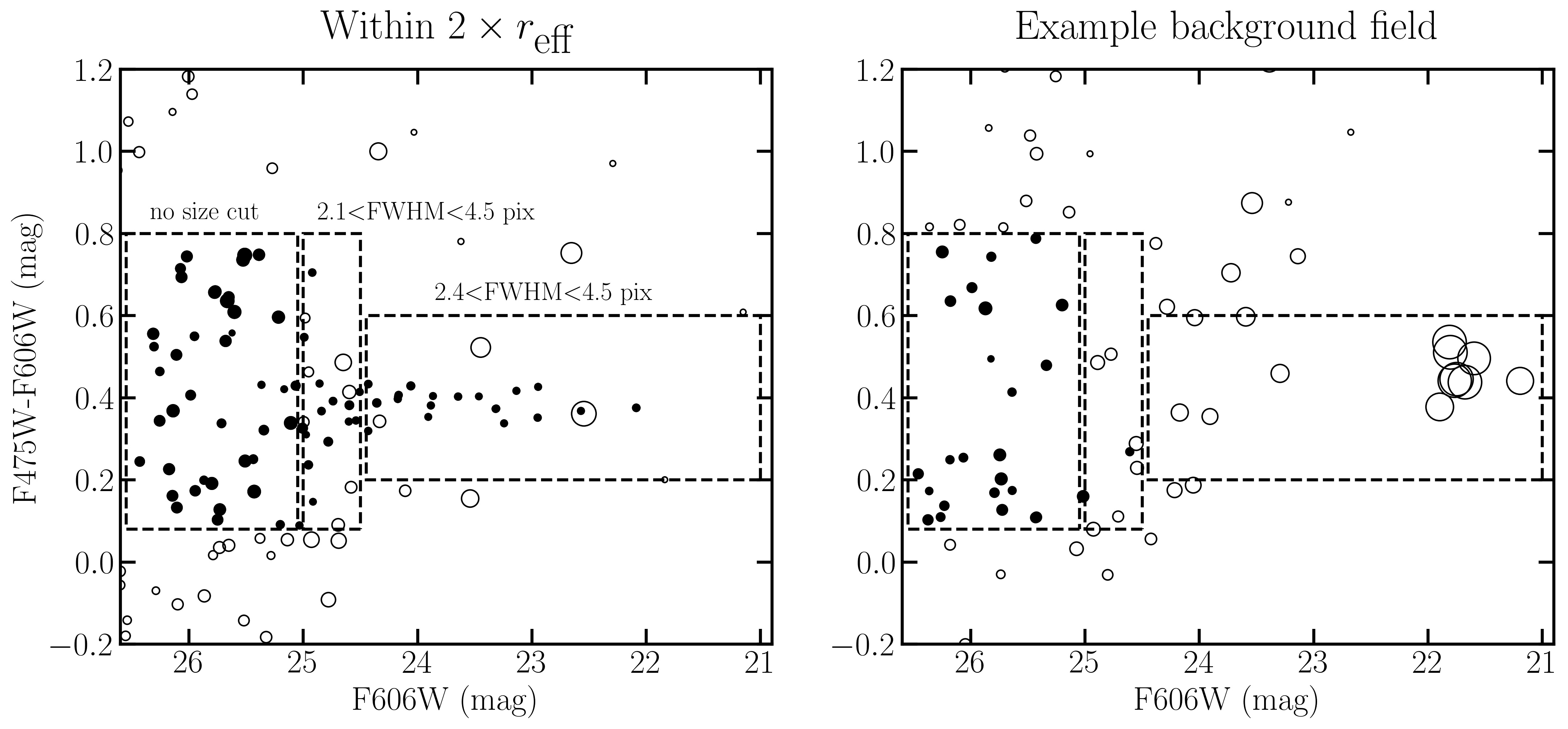}
  \caption{Photometric selection criteria for globular clusters in NGC5846-UDG1. Left panel: all sources within $2\times r_{\mathrm{eff}}$ in the color--magnitude plane. All sources have sizes proportional to their FWHM sizes measured in pixels. Sources that met the selection criteria are shown in the filled circles and open circles are sources that were not selected. The three dashed line boxes show the different color criteria as a function of the F606W magnitude. Right panel: same as the left panel for a background field with a similar area. There is only one contaminating source (filled circle) with $m_{\mathrm{F606W}}<25.0\,\mathrm{mag}$. For fainter sources ($m_{\mathrm{F606W}}>25.0\,\mathrm{mag}$) the number of contaminating objects increases, and these are accounted for in the calculation of the total globular cluster number.}
  \label{fig:selection}
}
\end{figure*}

\subsection{Identification of Globular Clusters} \label{sec:identification}

Identification of the globular clusters in NGC5846-UDG1 is based on their appearance in the 2-orbit WFC3/UVIS images. We also utilize the information of the 11 spectroscopically-confirmed globular clusters as detailed below. 

First, \texttt{SExtractor} \citep{Bertin1996:sextractor} is run on the F606W and F475W drizzled images in dual image mode where detection is done on the F606W images, with a $2\sigma$ detection threshold and a minimum area of $6\,\mathrm{pixels}$ above the threshold. The photometry is corrected for Galactic extinction \citep{Schlafly2011:reddening}. Total magnitudes were determined using the ``AUTO'' fluxes from \texttt{SExtractor}, corrected to an infinite aperture using the up-to-date UVIS2 encircled energy (EE) corrections\footnote{\url{https://www.stsci.edu/hst/instrumentation/wfc3/data-analysis/photometric-calibration/uvis-encircled-energy}}. All magnitudes are given in the AB system. The F475W$-$F606W color correction is negligible and therefore no correction was required for the wavelength dependence of the PSF. 
`
Similar to past studies \citep{vD2018:df2gcs, Shen2021:gcs}, globular cluster selection is done from the {\it HST} photometry using a set of size and color criteria. The selection is informed by the color and size distributions of the 11 spectroscopically-confirmed globular clusters with $(\mu, \sigma)_{\mathrm{F475W-F606W}}=(0.39, 0.03)\,\mathrm{mag}$, and $(\mu, \sigma)_{\mathrm{FWHM}}=(2.9, 0.29)\,\mathrm{pixels}$. We apply varying color and size criteria for different magnitude-selected sources to account for the variation in the photometric uncertainties with magnitude.  

Similar to \citet{vD2018:df2gcs} and \citet{Shen2021:gcs}, we create three size and color-selected source catalogs as demonstrated in Figure \ref{fig:selection}. We measure the half-light radius of the galaxy ($r_{\mathrm{eff}}$, see Section \ref{sec:structural}) and use the same selection criteria when generating source catalogs for the galaxy (within $2\times r_{\mathrm{eff}}$) and for a background field (outside $3\times r_{\mathrm{eff}}$). The background field is used for correcting for contaminating background and foreground sources. The first catalog includes sources brighter than $m_{\mathrm{F606W}}=24.5\,\mathrm{mag}$ and within a narrow color and size range of $0.2<\mathrm{F475W-F606W}<0.6\,\mathrm{mag}$, and $2.4<\mathrm{FWHM}<4.5\,\mathrm{pixels}$, respectively (dashed purple box). The same selection is made on sources that are outside $3\times r_{\mathrm{eff}}$ for background subtraction. Using these criteria we find $20$ sources within $2\times r_{\mathrm{eff}}$ and two sources outside $3\times r_{\mathrm{eff}}$. Accounting for the relative area differences ($A_{\mathrm{bg}} = 6.7\times A_{\mathrm{gal}}$), this first catalog has a background contamination of $0.3$ sources within $2\times r_{\mathrm{eff}}$. We repeat the same procedure for fainter sources using two more selection criteria. For sources with $24.5<m_{\mathrm{F606W}}<25.0\,\mathrm{mag}$ we allow a wider color range criterion of $0.08<\mathrm{F475W-F606W}<0.8\,\mathrm{mag}$ and $2.1<\mathrm{FWHM}<4.5\,\mathrm{pixels}$ (dashed green box). We identity $13$ sources within $2 \times r_{\mathrm{eff}}$ and $0.7$ background contaminants. Finally, for sources with $25.0<m_{\mathrm{F606W}}<26.5\,\mathrm{mag}$, we apply the same color cut of $0.08<\mathrm{F475W-F606W}<0.8\,\mathrm{mag}$ and no size cut (dashed orange box). The final selection adds a total of $43$ sources inside $2\times r_{\mathrm{eff}}$ and $24.5$ background contaminants. The left panel of Figure \ref{fig:selection} shows all sources within $2\times r_{\mathrm{eff}}$ from the center of the galaxy, color-coded and with circle sizes proportional to their FWHM sizes measured with \texttt{SExtractor}. Open circles represent sources that were not selected and filled circles are those sources that meet both the color and size criteria, for their F606W magnitude. The right panel shows the same selection applied to an example background field. The contamination for the bright sources catalog is very low, with essentially no contaminating sources below $m_{F606W}<24.5\,\mathrm{mag}$.  

In total, we find a background-corrected number of 50.4 globular clusters within $2\times r_{\mathrm{eff}}$. Assuming a \sersic\ distribution of the globular clusters with the same \sersic\ index for the smooth light component ($n=0.61$, see Section \ref{sec:structural}) and half-light radius, we estimate that three more globular clusters should reside outside $2 \times r_{\mathrm{eff}}$. We also correct for sources fainter than $26.5\,\mathrm{mag}$, assuming a Gaussian distribution, adopting the best-fit mean and scatter of $\mu=24.7\,\mathrm{mag}$ and $\sigma=1.1\,\mathrm{mag}$ (see Section \ref{sec:fraction}). Together, we obtain $53.9 \pm 8.9$ globular clusters associated with NGC5846-UDG1. The uncertainty in the total globular cluster count is dominated by the number of background sources at the faint end of the luminosity function and reflects Poisson errors in the observed counts (galaxy$-$background).

A lower globular cluster count in NGC5846-UDG1 has been previously determined using ground-based data ($\sim 45$, no error bars provided; \citealt{Forbes2021:kcwi}) and also using slightly shallower, single-orbit {\it HST} ACS data \citep{Muller2021:HST}. These authors report $26 \pm 6$ globular clusters, compared to $54 \pm 9$ found here. With the information provided in \citet{Muller2021:HST} we were not able to perform a direct comparison with our analysis. We suspect that the difference in depth (two versus one orbit) and sampling (0\farcs04/pixel with WFC3 compared to 0\farcs05/pixel with ACS) between the datasets may be indirectly responsible for the difference. Owing to the brighter photometric limits \citet{Muller2021:HST} used quite restrictive criteria, such that the smallest and faintest sources are likely excluded from their analysis.

\section{Structural and Physical Parameters} \label{sec:structural}

\begin{figure*}[t!]
{\centering
  \includegraphics[width=1.0\textwidth]{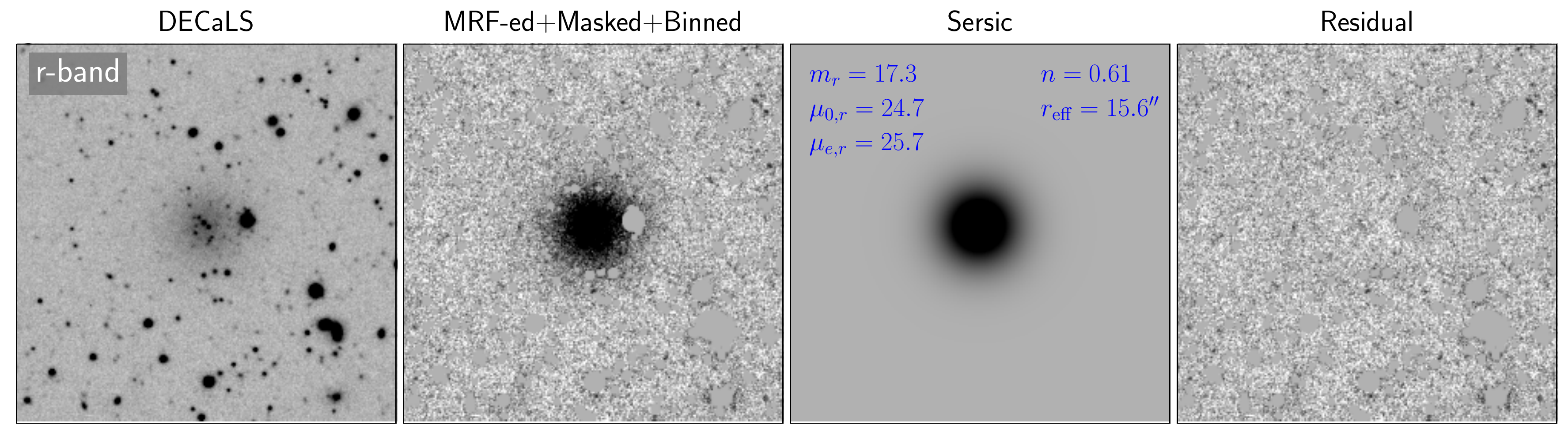}
  \caption{Left panel: $47'' \times 47''$ DECaLS $r$-band cutout centered on NGC5846-UDG1. Second to left panel: the same cutout after multi-resolution filtering. High surface brightness and compact sources were subtracted and masked. Second to right panel: two-dimensional \sersic\  best-fit model obtained using the \texttt{pymfit} code. The best-fit model parameters are shown in the blue text. Right panel: the residual image after subtracting the galaxy model.}
  \label{fig:mrf}
}
\end{figure*}

NGC5846-UDG1 is located in the NGC5846 galaxy group, at a projected distance of $21'$ from NGC5846 itself. It has a radial velocity of $2167\pm2\,\mathrm{km~s^{-1}}$ \citep{Forbes2021:kcwi}, consistent with it being a member of the group, which has a mean velocity of $1828\,\mathrm{km~s^{-1}}$ and a velocity dispersion of $\sim 295\,\mathrm{km~s^{-1}}$ \citep{Muller2020:muse}. We use the $i$-band image from the Hyper Suprime-Cam Subaru Strategic Program PDR2 (HSC-SSP; \citealt{Aihara2019:hsc}) to measure the surface brightness fluctuations (SBF) distance to NGC5846-UDG1. The SBF signal is measured using techniques developed for dwarf galaxies, with a calibration of $\overline{\mathrm{M}}_i=-1.29 \pm 0.22\,\mathrm{mag}$ based on the color of NGC5846-UDG1, as described in \citet{Carlsten2019:SBF} $\&$ \citet{Greco2021:SBF}. We measure SBF magnitude of $\overline{\mathrm{M}}_i^{\mathrm{UDG1}}=30.3 \pm 0.5 \,\mathrm{mag}$, which places NGC5846-UDG1 at $D\sim 21\pm5$ Mpc. This is consistent with the distance to the NGC5846 group of $26.5 \pm 0.8\,\mathrm{Mpc}$, reported in \citet{KourkchiTully2017:dis}, based on the weighted averaged distance of any members available in the Cosmicflows-3 distance catalog \citep{Tully2016:CF3}. We adopt $D=26.5\,\mathrm{Mpc}$ as the distance to NGC5846-UDG1 due to the low signal-to-noise achieved using the SBF technique, which places it at a projected distance of $162\,\mathrm{kpc}$ from NGC5846. We note that the main result of the paper, namely the globular cluster-to-field stars fraction, as shown below, is very insensitive to the exact distance to NGC5846-UDG1. 

Using {\it HST} and DECaLS data in tandem, we characterize the galaxy's physical and structural properties. The size, luminosity, surface brightness, and stellar mass of NGC5846-UDG1 were derived using the DECaLS images. Attempts were made to use the {\it HST} images for this purpose but the galaxy's surface brightness is too low for reliably fitting it with a model image. Instead, we utilized the {\it HST} images in a different way. We used the multi-resolution filtering (MRF) software \citep{vD2020:mrf} to remove all compact sources from the DECaLS images and isolated the diffuse component of NGC5846-UDG1. Briefly, \texttt{SExtractor} is run on the high resolution  {\it HST} images, and the resulting segmentation map was converted to a mask and multiplied by the image to create a flux model of all detected sources. Low surface brightness objects and saturated stars were removed from this model. The model was then convolved with a kernel to match the DECaLS point spread function and subtracted. The original $r$-band DECaLS and final residual image binned $2 \times 2$, conserving only the low surface brightness component, are shown in the left two panels of Figure \ref{fig:mrf}.

Next, we parameterized the galaxy's structure by fitting a two-dimensional \sersic\ model to the smooth light component in the DECaLS $g$ and $r$-band images. We used the \texttt{pymfit}\footnote{\url{https://github.com/johnnygreco/pymfit}} code, a python wrapper of \texttt{imfit} \citep{Erwin2015:imfit}, on the MRFed, masked, and binned ($2 \times 2$) DECaLS images. First, \texttt{imfit} was run on the $r$-band MRFed image to determine the structural parameters of the galaxy and its brightness. Then the $g$-band MRFed image was fitted while allowing only the amplitude to change. To estimate the galaxy fit uncertainties, we injected the best-fit \texttt{imfit} model into different areas in the MRFed and binned $g$ and $r$ images, and then fitted the injected models in the same way we fitted the galaxy. We adopt the RMS variation in the fitted parameters as the uncertainty for each parameter. The final model and residual $r$-band image are shown in the two right panels of Figure \ref{fig:mrf}. 

The best-fit model has a \sersic\ index of $n=0.6$, axis ratio $b/a=1.0$, and a central surface brightness of $\mu_{g,0} = 25.4\,\mathrm{mag\,arcsec}^{-2}$. It has a half-light radius of $r_{\mathrm{eff}}=15.6''$, corresponding to $r_{\mathrm{eff}}=1.9\,\mathrm{kpc}$ at $26.5\,\mathrm{Mpc}$. We transform the $g$ and $r$-band measured quantities into $V$-band magnitude and luminosity\footnote{\url{http://www.sdss3.org/dr8/algorithms/sdssUBVRITransform.php\#Lupton2005}}. Its total $V$-band absolute magnitude is $M_V=-14.6\,\mathrm{mag}$ and its luminosity is $L_V = 5.9 \times 10^7\,\mathrm{L_{\odot}}$. With a mass-to-light ratio $M/L_V=2.0$, based on $[\mathrm{Fe/H}] = -1.33$ and $\mathrm{age}=11.2\,\mathrm{Gyr}$ from \citet{Muller2020:muse}, we obtain a stellar mass of $M_\star \approx 1.2 \times 10^8\,\mathrm{M}_\odot$.

\section{Fraction of Stars in Globular Clusters} \label{sec:fraction}

\begin{figure*}
\centering
\includegraphics[width=0.95\textwidth]{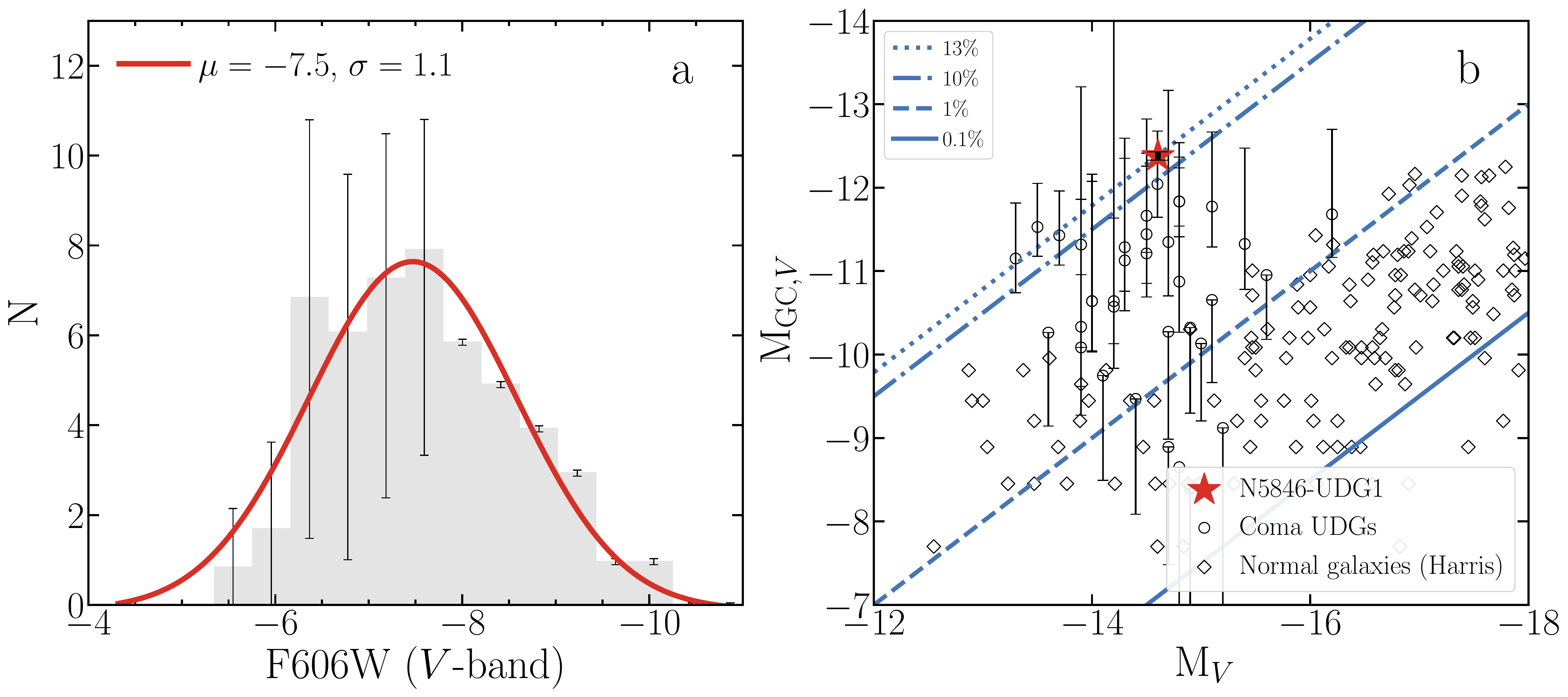}\label{fig:lf}
\caption{Left panel: the background-subtracted globular cluster luminosity function of NGC5846-UDG1, in absolute magnitude (gray histogram), and its best-fit Gaussian luminosity function (red curve). Right panel: the absolute magnitude in globular clusters vs. the total magnitude of the host galaxy. Diagonal lines show constant globular cluster-to-field stars light fractions, ranging from $0.1\%$ to $13\%$. NGC5846-UDG1, shown as a red star, has the highest fraction of stars in globular clusters known to date ($12.9 \pm 0.6\%$). This is measured with high certainty, demonstrated with the small error bar ($0.05\,\mathrm{mag}$) on top of the star. Some Coma ultra-diffuse galaxies (empty circles) are excellent candidates for sharing a similar high fraction of stars in globular clusters, but these are four times farther away and thus 16 times fainter and have commensurately larger error bars.}
\end{figure*}

In the left panel of Figure \ref{fig:lf} we show the globular cluster luminosity function and a best-fit Gaussian function. Assuming a distance of $26.5\,\mathrm{Mpc}$, it peaks at $M_V=-7.5\,\mathrm{mag}$ with a width of $\sigma=1.1\,\mathrm{mag}$, consistent with canonical values for globular cluster systems in most other galaxies \citep{Rejkuba2012:gclf}. 

We calculate the fraction of light in globular clusters relative to the total light of the galaxy. The total flux in globular clusters was obtained by integrating the F606W and F475W histograms (scaled to account for the background subtraction). We calculate the $1\sigma$ and $3\sigma$ uncertainties of the integrated globular cluster magnitude by generating 100,000 realizations of the histograms and calculating the $1\sigma$ ($16\%, 84\%$) and $3\sigma$ ($0.15\%, 99.8\%$) intervals of the total magnitude distribution. Each histogram is created by perturbing the measured histogram values according to the error bars in each bin, for magnitude bins fainter than $M=-7.5\,\mathrm{mag}$. The total light in the identified globular clusters is dominated by the most massive clusters and is therefore insensitive to the completeness limits. 

We find that the fraction of light in globular clusters is unusually high, given the galaxy total absolute magnitude, with $M_{V,\mathrm{GCs}}=-12.5 \pm 0.05 \,\mathrm{mag}$. This implies that $12.9 \pm 0.6\%$ of the stars (measured in $V$-band) currently reside in the globular clusters, a fraction that is about $100\times$ greater than for the Milky Way \citep{Harris2013:gcs}.

\section{Discussion}

\begin{figure*}[t!]
\centering
\includegraphics[width=1.0\textwidth]{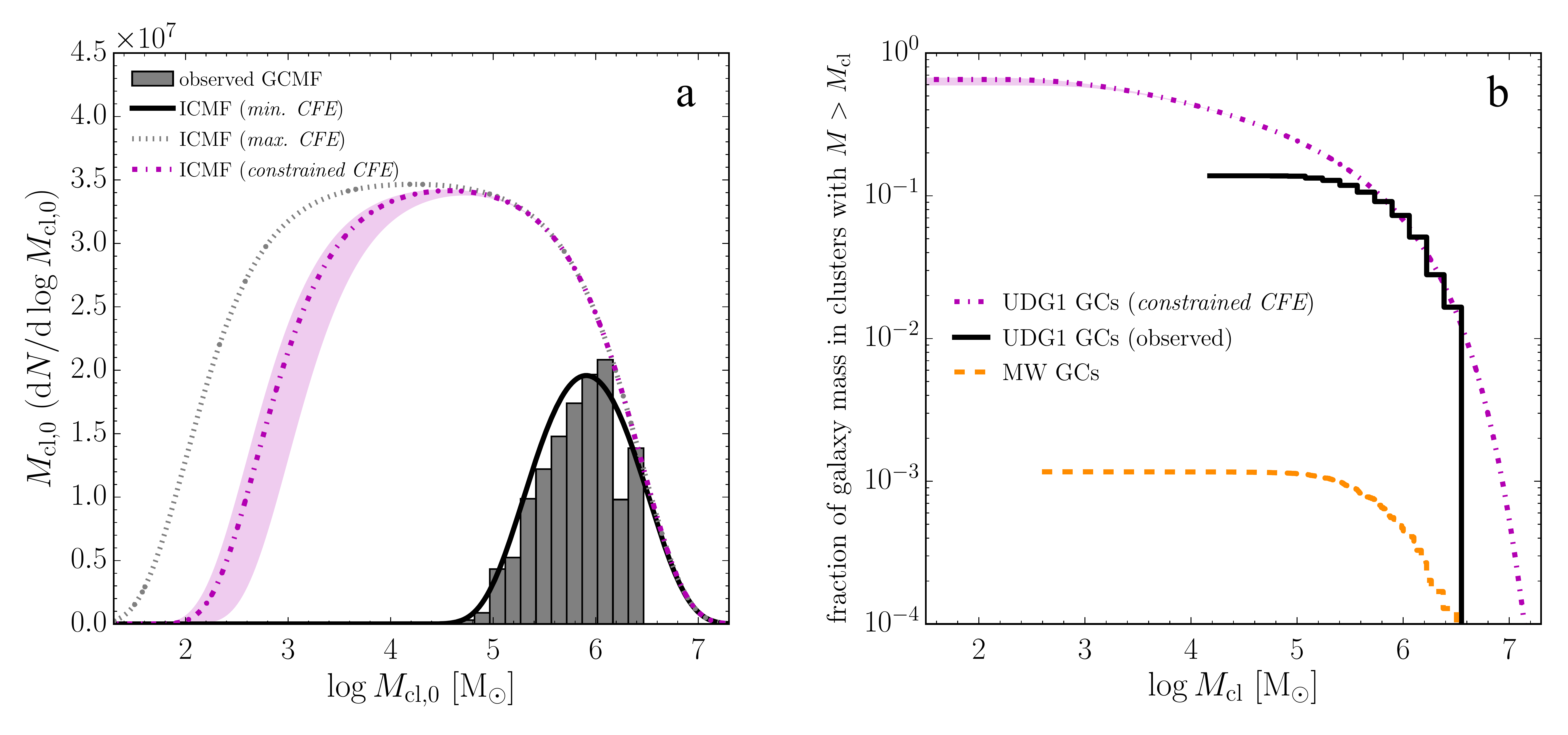}\label{fig:model}
\caption{Initial cluster mass function (ICMF) models constrain the origin of NGC5846-UDG1's globular cluster system. Left panel: constraints on the ICMF of NGC5846-UDG1, based on its present-day observed globular clusters and total stellar mass. The observed globular cluster mass function, corrected for mass loss due to stellar evolution, is shown in the black histogram. The \textit{fiducial model}, constrained by the environmental dependence of the fraction of star formation in bound clusters (i.e. the CFE; \citealt{Kruijssen2012:sf}), is shown as the shaded purple band for a range of Toomre $Q$ disk stability parameters ($0.5 \leq Q \leq 3.0$), where the purple dot-dashed curve corresponds to $Q=0.5$. For reference, the black solid curve and the dashed gray curve show the \textit{minimum} and \textit{maximum CFE models}, respectively. The \textit{minimum CFE model} assumes that the present-day globular clusters were the only clusters that formed, whereas the maximum CFE model assumes that all stars in the galaxy came from disrupted low-mass clusters. The mass functions are weighted by mass, so that the area under each curve is directly proportional to the mass in clusters. Right panel: the cumulative fraction of galaxy mass in clusters at the time of formation for NGC5846-UDG1 in comparison to the Milky Way. For NGC5846-UDG1, the purple dot-dashed curve shows the \textit{fiducial model} (constrained by the CFE) and the black solid curve shows the \textit{minimum CFE model}. The orange dashed curve assumes no mass loss or cluster disruption for the Milky-Way globular clusters, correcting only for mass loss due to stellar evolution. The fraction of galaxy mass in clusters in NGC5846-UDG1 is 2--3 orders of magnitude higher than in the Milky Way, irrespectively of the adopted \textit{CFE model}.}
\end{figure*}

\subsection{Modeling of the Cluster-to-Field Stars Mass Ratio}

NGC5846-UDG1 is the first known galaxy with a present-day globular cluster-to-field stars mass ratio that is $\geq 10\%$ with $>90$\,\% confidence. Fractions that are this high have important implications for formation models of the galaxy, while providing new constraints on globular cluster mass loss. On long enough time scales, clusters across all masses evolve dynamically, including the complete dissolution of low-mass clusters. These processes include tidal shocks through interactions with the substructure of the dense interstellar medium and evaporation as a result of two-body relaxation \citep{Krause2020:clusterformation}. Taking into account complete destruction of low mass coeval proto-globular clusters as well as mass loss from the surviving clusters, it is thought that initially the cluster population contained up to a factor of a few more stars than are observed after $\sim 10$\,Gyr. That is, globular cluster systems could lose up to $\sim 80-90\%$ of their stars over their lifetime \citep{Larsen2012, Kruijssen2015, Reina-Campos2018}. With a measured fraction of $12.9 \pm 0.6\%$, NGC5846-UDG1 could have started with a globular cluster-to-field stars mass fraction approaching $100\%$. Below we quantify this result using the specific properties of NGC5846-UDG1.

We apply an analytical physical model for the globular cluster mass function (GCMF) to infer the environmental conditions within the progenitor galaxy at the time of formation of the globular clusters \citep{Reina-Campos2017, TG2019:model}. The model relates the properties of the host galaxy disk at the time of formation (its gas surface density, rotational angular velocity, and Toomre $Q$ stability parameter) to the minimum and maximum cluster masses \citep{Reina-Campos2017, TG2019:model}, as well as the cluster formation efficiency (CFE; the fraction of star formation occurring in gravitationally bound clusters; \citealt{Kruijssen2012:sf}). Together with the assumption of a power-law slope of $-2$ between the minimum and maximum exponential truncation masses, these quantities completely determine the double-Schechter initial cluster mass function (ICMF) \citep{TG2019:model, Adamo2020:HiPEEC}. We assume $M/L_V \approx 2.0$, and correct for the fractional mass loss due to stellar evolution of $f=0.35$ for the field stars and $f=0.40$ for the globular clusters, to also account for mass segregation \citep{Lamers2010}. This model accurately reproduces the demographics of young and old cluster populations in the local Universe \citep{Pfeffer2019, TG2019:model}. 

Since we do not know a priori if stars that are currently not in globular clusters were formed as part of lower-mass proto-globular clusters or as field stars, we model the ICMF making three distinct assumptions that bracket the range of possibilities. The left panel of Figure \ref{fig:model} shows the ICMF for the three models which account for the inferred fraction of the total stars formed in bound clusters. We first consider the two extreme cases. The black curve shows the \textit{minimum CFE model} where only the stars currently observed in clusters were born in gravitationally bound clusters, and the ICMF is simply the present-day GCMF corrected for mass loss due to stellar evolution (i.e., with no dynamical mass loss). Second, the gray dashed curve shows the \textit{maximum CFE model} where all the stars in the galaxy were born in bound clusters, and assuming that the integrated ICMF mass cannot exceed the total stellar mass of the galaxy. Lastly, using the purple band we show the \textit{fiducial 'constrained CFE' model}, which uses the additional requirement that the predicted set of environmental properties simultaneously reproduces the observed minimum and maximum masses, and the predicted fraction of star formation in bound clusters. The \textit{fiducial model} is shown for a disk stability range of $0.5 \leq Q \leq 3.0$.

In the model that is preferred for describing a rapidly cluster-forming galaxy (with $Q=0.5$, which reflects the typical values observed in high-redshift galaxies; \citealt{Genzel2014}), we obtain a minimum cluster mass of $M_{\mathrm{min}}=4.9\times 10^2\,\mathrm{M}_\odot$ and a bound fraction of $65 \pm 2\%$. This implies that to satisfy the interstellar medium (ISM) conditions required to produce the observed GCMF (gas surface density of $\Sigma_{\mathrm{ISM}}=4.2 \pm 0.6 \times 10^2 \mathrm{M}_{\odot}\,\mathrm{pc}^{-2}$ and shear $\Omega = 0.24^{+0.18}_{-0.10}\,\mathrm{Myr}^{-1}$), a very large fraction of the total mass of NGC5846-UDG1 had to form in bound clusters. We therefore infer that it is likely that most of the star formation in NGC5846-UDG1 happened in extremely dense cluster-forming gas clumps (in order to produce clusters that remain bound after losing their gas), with unusually little low-density (i.e. unbound) star formation occurring \footnote{The model assumes that the majority of field stars and clusters formed during the same star formation episode.}. 

The right panel of Figure \ref{fig:model} shows the cumulative fraction of galaxy mass in clusters, at formation, for NGC5846-UDG1 in comparison to the Milky Way (orange dashed curve), assuming no cluster mass loss or disruption for the Milky Way (i.e. the \textit{minimum CFE model}): $M/L_V = 2.0$ is assumed for the Milky Way globular clusters \cite{Harris1996} and a stellar mass of $M_{\star}^\mathrm{MW}=5 \times 10^{10}\,\mathrm{M_{\odot}}$ \citep{BH-Gerhard2016:Galaxy}. The black solid curve is the \textit{minimum CFE model} for NGC5846-UDG1 while the purple dot-dashed curve shows the \textit{fiducial model}. There is a 2--3 orders of magnitude difference between NGC5846-UDG1 and the Milky Way, further demonstrating the extraordinary conditions that led to the formation of NGC5846-UDG1. While formation in massive, gravitationally bound clusters made only minor contributions to the total Milky Way mass, star formation in dense clumps is the dominant formation channel in NGC5846-UDG1.  

\subsection{Tests of the Model}

How can this scenario be further tested observationally? If indeed a large fraction of the field stars came from dissolved clusters, then the expectation is that both the stellar populations (particularly the ages) and the spatial distribution of the field stars and the globular clusters are similar. In other galaxies, this is typically not the case, with globular clusters on average being more metal-poor and older than field stars, and their spatial distribution more extended with $R_{\mathrm{eff,GC}} \approx 1.5-2 R_{\mathrm{eff},\star}$ \citep{Kartha2014:sluggs, vD2017:gcs, Lim2018:gcs}. Recent spectroscopic data have shown that the globular clusters and the field stars of NGC5846-UDG1 have consistent ages (with $\mathrm{age}=11.2^{+1.8}_{-0.8}\,\mathrm{Gyr}$) and metallicities ($\mathrm{[Fe/H]}=-1.33^{+0.19}_{-0.01}$ for the field stars) \citep{Muller2020:muse}. We measure the half-number radius of the globular clusters by selecting all sources from the entire {\it HST} image that have $V$-band magnitudes brighter than $25\,\mathrm{mag}$ (fainter than the expected canonical peak at $24.6\,\mathrm{mag}$), that are relatively compact ($2.2<\mathrm{FWHM}<5.0\,\mathrm{pixels}$), and have $g-V$ colors between $0.2$ and $0.6$ mag. We identify 34 globular clusters in the sample, of which 31 are located very close to the galaxy. With this low-contamination sample ($N_\mathrm{contamination}=0.15$ clusters), we find $R_{\mathrm{eff,GC}}=12.6 \pm 1.8''$, consistent with the field star half-light radius of $R_{\mathrm{eff}}=15.6 \pm 0.8''$. The similar (and possibly more concentrated) distribution of globular clusters compared to the smooth stellar distribution, as well as the similar ages and chemical compositions, support a common origin of the stars in globular clusters and the field stars. Future studies may be able to refine the age estimates of the clusters and the diffuse light.

\subsection{NGC5846-UDG1 in Context}

NGC5846-UDG1 is likely not alone. Globular cluster-rich UDGs have been identified and studied in various cosmic environments \citep{vD2017:gcs, Beasley2016, Forbes2020:ComaUDGS, Somalwar2020}. In fact, the large populations of globular clusters in such low-luminosity and diffuse systems has been one of the properties that make the formation of UDGs stand out as perplexing in the context of modern physical models of galaxy formation. We put NGC5846-UDG1 in context with other globular cluster-rich galaxies with absolute magnitudes ranging between $M_V=-12\,\mathrm{mag}$ and $M_V=-18\,\mathrm{mag}$. In the right panel of Figure \ref{fig:lf}, we show the total $V$-band absolute magnitude of galaxies as a function of the total absolute magnitude of globular clusters in these galaxies. NGC5846-UDG1 has a larger confirmed fraction of luminosity in its globular cluster system than any of the other galaxies. Interestingly, some UDGs in the Coma cluster might possess similar extreme fractions of stars in their present-day globular cluster systems. However, the Coma Cluster is located $\sim 4$ times farther than NGC5846-UDG1 and correspondingly the uncertainties on estimates of potentially associated globular clusters are much larger. These uncertainties are dominated by the high contamination from background and foreground objects at fainter magnitudes and the fact that their globular clusters are unresolved with {\it HST} at $100\,\mathrm{Mpc}$. Nevertheless, these Coma galaxies suggest that clump-only star formation in galaxies may be quite common. 

What conditions and processes might have promoted star formation, predominating in dense gas clumps, leading to the formation of a galaxy such as NGC5846-UDG1? The ICMF model applied to NGC5846-UDG1 predicts high gas surface densities ($\Sigma_{\mathrm{ISM}}=4.2 \pm 0.6 \times 10^2 \mathrm{M}_{\odot}\,\mathrm{pc}^{-2}$) at the time of formation. A recent model that connects the evolution of galaxies with their dark matter halos and globular cluster populations predicts that high globular cluster mass fractions arise naturally in early-collapsing dark matter halos due to their elevated gas surface densities at high redshift \citep{TG2021:DF2}. Such conditions give rise to massive cluster formation with a high CFE, resulting in an elevated number of globular clusters relative to the galaxy stellar mass. The model then predicts that the high clustering of the supernova feedback sources within the clusters could significantly increase the mass loading of gas outflows, which would lead to significant expansion of the stars and dark matter compared to galaxies with typical halo collapse times and globular cluster populations. For NGC5846-UDG1, such a short-lived and efficient burst of star formation at high redshift may have also expelled the remaining gas, limiting star formation to this single event. The early-collapse UDG formation model predicts a correlation between offset from the mean stellar mass-halo mass relation and number of formed GCs. Precise constraints on the dark matter halo mass of NGC5846-UDG1 could therefore be used to test this scenario. NGC5846-UDG1 may be a case where star formation began like in many other galaxies, but then failed to form stars in a low-density mode at later epochs.

\acknowledgments
Support from STScI grant HST-GO-16284 is gratefully acknowledged. SD is supported by NASA through Hubble Fellowship grant HST-HF2-51454.001-A awarded by the Space Telescope Science Institute, which is operated by the Association of Universities for Research in Astronomy, Incorporated, under NASA contract NAS5-26555. STG and JMDK gratefully acknowledge funding from the European Research Council (ERC) under the European Union’s Horizon 2020 research and innovation programme via the ERC Starting Grant MUSTANG (grant agreement number 714907). JMDK gratefully acknowledges funding from the Deutsche Forschungsgemeinschaft (DFG) in the form of an Emmy Noether Research Group (grant number KR4801/1-1). STG gratefully acknowledges funding from the European Research Council (ERC) under the European Union’s Horizon 2020 research and innovation programme via the ERC Starting Grant MUSTANG (grant agreement number 714907). AJR was supported by National Science Foundation grant AST-1616710 and as a Research Corporation for Science Advancement Cottrell Scholar.


\end{document}